\documentclass[prl,twocolumn,showpacs,preprintnumbers,amsmath,amssymb]{revtex4}
\usepackage{graphicx}
\begin{document}
\title{Surface dangling bond states and band-lineups in 
hydrogen-terminated Si, Ge, and Ge/Si nanowires}
\author{R. Kagimura}
\author{R. W. Nunes}
\author{H. Chacham}\thanks{corresponding author}
\email{chacham@fisica.ufmg.br}
\affiliation{Departamento de F\'{\i}sica, ICEX, Universidade Federal de 
Minas Gerais, CP 702, 30123-970,Belo Horizonte, MG, Brazil.}

\date{\today}

\begin{abstract} 
We report an {\it ab initio} study of the electronic properties of
surface dangling-bond (SDB) states in hydrogen-terminated Si and Ge
nanowires with diameters between 1 and 2 nm, Ge/Si nanowire
heterostructures, and Si and Ge (111) surfaces.  We find that the
charge transition levels $\varepsilon(+/-)$ of SDB states behave as a
common energy reference among Si and Ge wires and Si/Ge
heterostructures, at 4.3 $\pm$ 0.1 eV below the vacuum
level. Calculations of $\varepsilon(+/-)$ for isolated atoms indicate
that this nearly constant value is a periodic-table atomic property.
\end{abstract}


\maketitle

When a junction of two distinct, undoped semiconductor materials (a
heterojunction) is formed, one can define the band-lineup using the
valence- and conduction-band edge energies on each side of the
junction. The measurement or calculation of such lineup can be made
either directly from the band discontinuities at the interface, or
indirectly through a common bulk energy reference on both sides of the
interface.  From the theoretical side, several such common energy
references have been proposed.  Tersoff suggests~\cite{tersoff} that
an effective midgap energy $E_B$ (which can be calculated from the
bulk electronic structure) can be used to predict lineups within an
accuracy of 0.2 eV. Electronic states of transition metal
impurities~\cite{ledebo,zunger1,tersoff2} and charge transition levels
$\varepsilon(+/-)$ of hydrogen interstitial impurities
~\cite{naturevw03,zunger2} have also been shown to work as common
energy references.

In the present work, we report an {\it ab initio} study of the
electronic properties of surface dangling bond (SDB) states in
hydrogen-terminated Si and Ge thin nanowires (with diameters between 1
and 2 nm), Ge/Si nanowire heterostructures, and Si and Ge (111)
surfaces. Here we define the SDB as the defect resulting from the
incomplete passivation of a Si or Ge surface atom, due to a missing
hydrogen.  These defects are observed in hydrogenated Si surfaces, and
can be used as ``gates" in molecular electronics devices~\cite{piva}.
They are also expected to occur in the recently synthesized ultra-thin
hydrogen-terminated Si nanowires~\cite{science03}. We find that the
charge transition levels $\varepsilon(+/-)$ of SDB states are deep in
the bandgap for the Si surface and wires, shallow for the Ge wires,
and resonant in the valence band for the Ge (111) surface. We also
find that the SDB $\varepsilon(+/-)$ levels behave as a common energy
reference level among Si and Ge wires and Ge/Si heterostructures, at
4.3 $\pm$ 0.1 eV below the vacuum level. Calculations of
$\varepsilon(+/-)$ for Si and Ge isolated atoms, as well as for group
III and V atoms, indicate that this constant value is a periodic-table
atomic property.

Our calculations are performed in the framework of Kohn-Sham density
functional theory (DFT)~\cite{ks}, within the generalized-gradient
approximation (GGA)~\cite{gga} and norm-conserving
pseudopotentials~\cite{pse-tm} in the Kleinman-Bylander factorized
form~\cite{KL}. We use the LCAO method implemented in the SIESTA
code~\cite{siesta}, with a double-zeta basis set plus polarization
orbitals. Total-energy differences (forces) are converged to
within 10~meV/atom (40~meV/\AA) with respect to calculational
parameters, that are the same as those of our previous
work~\cite{kagimura}.

We consider nanowires based on the diamond bulk phase, oriented along
the (110), (111), and (112) directions, with one SDB per unit cell.
The H-terminated nanowire structures are obtained by choosing the
nanowire axis along a given crystalline direction in the bulk lattice,
and including all atoms that fall within a specific distance from the
axis. The lowest-coordinated surface atoms are removed, and the
dangling bonds of the remaining surface atoms are saturated with
hydrogen. Depending on the nanowire diameter and orientation, three
types of H passivation are possible, related to whether the surface Si
(Ge) is bonded to one, two, or three H atoms, forming SiH (GeH),
SiH$_2$ (GeH$_2$), or SiH$_3$ (GeH$_3$) radicals on the surface,
respectively.

\begin{figure}[t]
\includegraphics[height=7.cm]{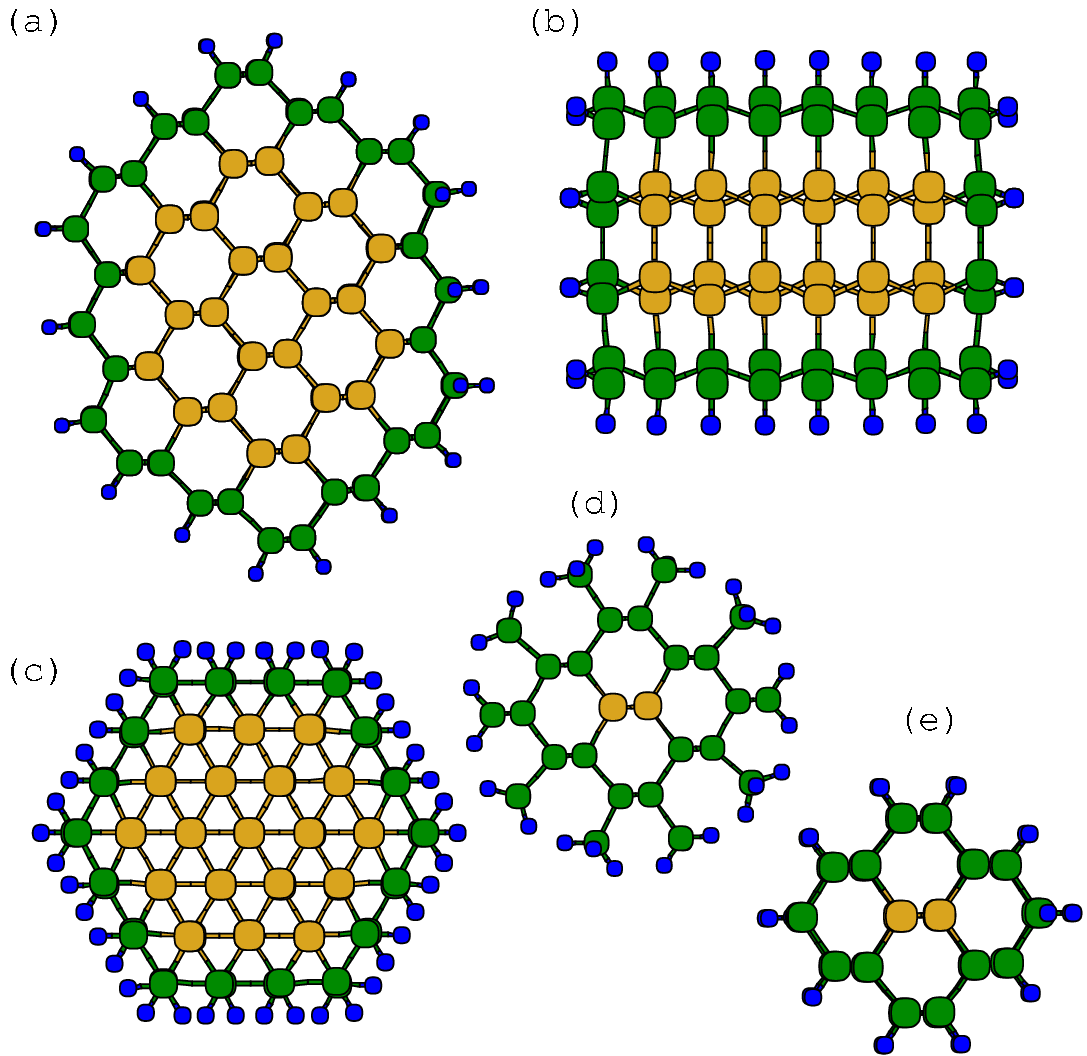}
\caption{(Color online) Cross section of several Ge/Si coaxial
nanowire heterostructures.  In (a), (d), and (e), nanowires with the
axis along the (110) direction and with diameters ($D$) of 23 \AA, 15
\AA\ and 12 \AA, respectively; in (b), nanowires with the axis along
the (112) direction and $D=19$ \AA; in (c), nanowires with the axis
along the (111) direction and $D=17$ \AA.  In all cases, the nanowire
axis is perpendicular to the plane of the paper. Si, Ge, and H atoms
are represented by green, yellow, and blue circles, respectively.}
\label{fiospas}
\end{figure}

\begin{figure}[t]
\includegraphics[height=3.8cm]{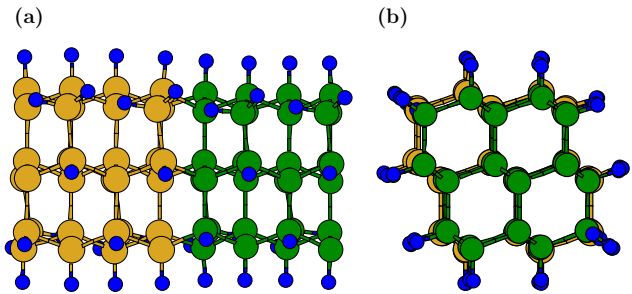}
\caption{(Color online) Ge/Si nanowire heterostructure with axis
oriented along the (110) direction and $D = 1.2~$nm.  Side (a)
and top (b) views of the unit cell are shown.}
\label{axial}
\end{figure}

Besides ``pure" Si and Ge nanowires, we also consider: coaxial Ge/Si
nanowires with a Ge core and a Si overlayer; a nanowire with a
periodic alternation of a few Ge and Si layers along the wire axis,
forming a Ge/Si nanowire heterostructure; and three-layer-thin Si and
Ge films, hydrogen-terminated on both sides, representing H-terminated
(111) surfaces.  The structures of the coaxial Ge/Si nanowires are
shown in Fig.~\ref{fiospas}. The structures of the pure Si and Ge
nanowires have the same initial geometry as those in
Fig.~\ref{fiospas}, with all Ge and Si atoms replaced by either
silicon or germanium. The geometry of the Ge/Si nanowire
heterostructure is shown in Fig.~\ref{axial}, with a periodic unit
along the axis formed by the alternation of four layers of Ge atoms
and four layers of Si atoms.  The atomic positions and lattice
parameter of all nanowires and (111) surfaces were relaxed within the
tolerances indicate above.

\begin{figure}[h]
\includegraphics[height=4.5cm]{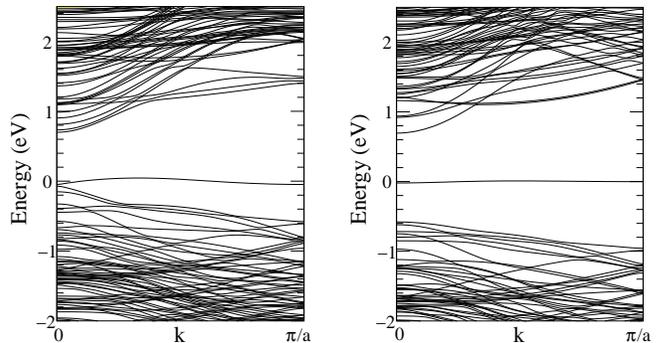}
\caption{Electronic band structures of Ge (left side) and Si (right
side) nanowires ($D \approx $2 nm) with one surface
dangling bond defect per unit cell. Both wires are oriented along the
(110) direction.}
\label{eeletronica}
\end{figure}

We now address the electronic structure of nanowires with one SDB
defect per unit cell, obtained by removing a hydrogen atom from the
nanowire surface and performing full {\it ab initio} structural
optimizations. We ensure that the unit cells we use are large enough
to decouple neighboring defects.  For the majority of the nanowires,
we consider SDB's generated from GeH and SiH radicals, with the
exception of wires with $D = 1.4~(1.5)~{\rm nm}$ for which SiH$_2$
(GeH$_2$) radicals are considered [no SiH (GeH) radicals are present
in this case].  Generally, we find that SDB Kohn-Sham eigenvalues are
deep in the bandgap in the case of Si wires, and shallow (near the
valence band edge) in the case of Ge wires.  This is seen in
Fig.~\ref{eeletronica}, that shows the calculated band structures for
the defective Si and Ge nanowires with $D \approx 2$~nm, oriented
along the (110) direction.  The unit cells are 7.9~\AA\ and 8.4~\AA\
long along the axis, for the Si and Ge wires, respectively. Both band
structures in Fig.~\ref{eeletronica} are essentially identical to
those of the perfect wires (not shown), except for the appearance of a
half-occupied nearly-dispersionless band in the gap. The small
dispersions (0.03 eV for Si and 0.09 eV for Ge) are characteristics of
localized SDB states.  These features of the SDB level are common to
all ten Si and Ge ``pure" nanowires we consider.

\begin{figure}[h]
\includegraphics[height=4.0cm]{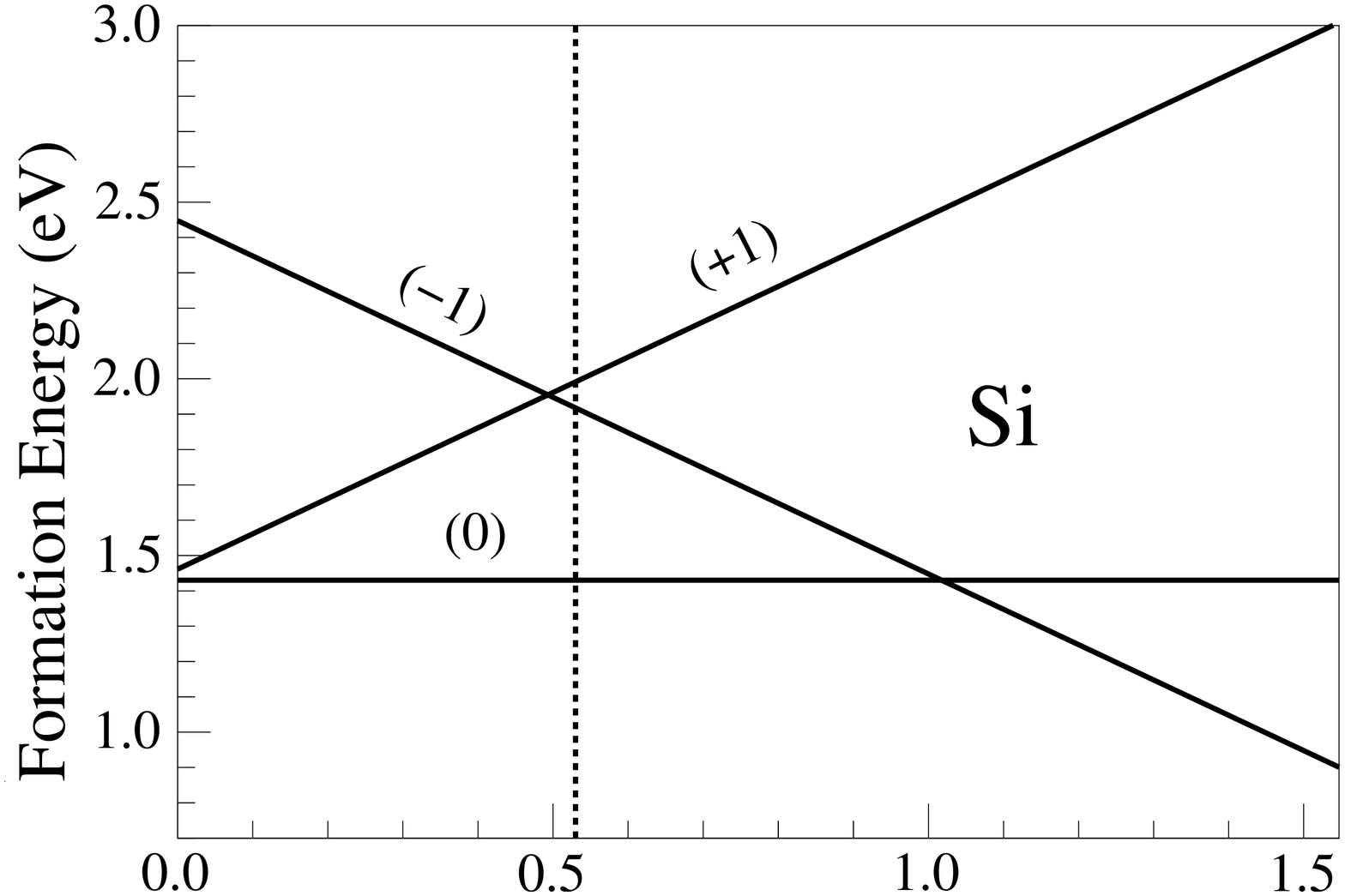}
\includegraphics[height=4.16cm]{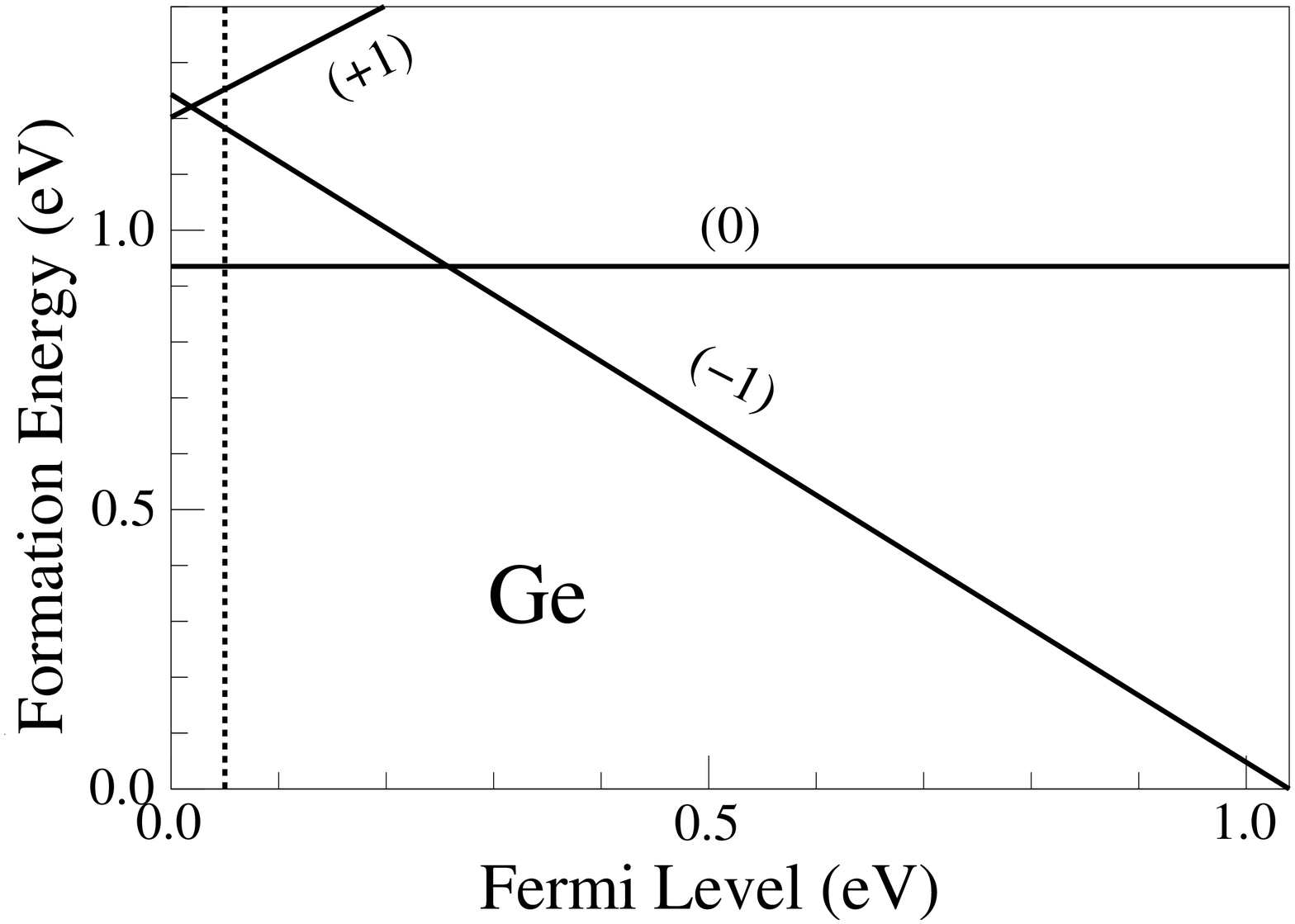}
\caption{
Formation energies of SDBs in a (112) Si wire (diameter of
19 \AA) and a (112) Ge wire (diameter of 20 \AA). The vertical dashed
line indicates the Kohn-Sham eigenvalue of the neutral SDB states.}
\label{fig.3}
\end{figure}

From Janak's theorem, $\varepsilon_i=\partial E/\partial
n_i$~\cite{janak}, and the nearly linear dependence of the SDB
eigenvalue $\varepsilon_i$ on its occupation number $n_i$, it follows
that the eigenvalue of the neutral SDB state at the GGA,
non-spin-polarized level, is a good approximation to the
$\varepsilon(+/-)$ charge transition level~\cite{footnotejanak}.  The
latter is given by the value of the electronic chemical potential at
which +1 and -1 charge states of a defect have equal formation
energies.  In Fig.~\ref{fig.3} we show a comparison between the SDB
neutral eigenvalue and the $\varepsilon(+/-)$ level, for the Si ($D=19
$ \AA) and Ge ($D= 20 $ \AA) (112) wires. The straight lines for each
charge state in Fig.~\ref{fig.3} are $E^{q}_{form}\left({\rm
SDB}\right) = E^{q}_{tot}\left({\rm SDB}\right) - E_{tot}({\rm
perfect}) + \mu_{\rm H} + q\mu $, where $E^{q}_{tot}\left({\rm
SDB}\right)$ is the supercell total energy for a nanowire with an SDB
in charge state $q$, $\mu$ is the Fermi level, and $E_{tot}({\rm
perfect})$ is the total energy for the perfect wire. The chemical
potential for hydrogen, $\mu_{\rm H}$, is obtained from a Si (Ge) bulk
and SiH$_4$ (GeH$_4$) molecule calculations.  The eigenvalue of the
neutral SDB state is plotted as a dashed vertical line, and is within
0.05 eV from the $\varepsilon(+/-)$ level for both Si and Ge.  We
obtained very similar results for Si and Ge wires oriented along the
(110) direction.  In the following, we present strong numerical
arguments showing that the $\varepsilon(+/-)$ level of nanowire SDB's
is a common reference level among Si, Ge, and Ge/Si nanowires.

Let us first consider the electronic structure of the Ge/Si coaxial
wires of Fig.~\ref{fiospas} with an SDB state at the surface.
Table~\ref{sige} indicates the positions of the neutral SDB level
relative to the valence band maximum, $E_d-E_v$, and relative to the
vacuum level, $E_d$. The table shows that $E_d-E_v$ varies from a
``shallow" value of 0.13 eV to a ``deep" value of 0.50 eV, although
the SDB's are, in all cases, at the surface Si atoms of the coaxial
wires.  In contrast, $E_d$ is a common energy reference for all the
SiH-terminated wires (i.e., all in Table~\ref{sige} except for the
wire with $D=1.5$ nm, which is SiH$_2$-terminated).

\begin{table}[!h]
\centering
\caption{Axis orientation, diameter, calculated bandgap, and
defect level position ($E_d$) [absolute and
relative to the valence band maximum ($E_v$)]
of Ge/Si coaxial nanowire heterostructures with one surface dangling bond
per unit cell. Energies are given in eV.}
\vspace{0.30cm}
\label{sige}
\begin{tabular}{lccccc} \hline \hline
   direction                   &
\multicolumn{1}{l}{(110)}&
\multicolumn{1}{l}{(112)}&
\multicolumn{1}{l}{(111)}&
\multicolumn{1}{l}{(110)}&
\multicolumn{1}{l}{(110)}\\
\hline
Diameter \:\:& 23 \AA \:\: & 19 \AA \:\: & 17 \AA \:\: & 15 \AA \:\:& 12 \AA \\
\hline
$\Delta E_{gap}$ \:\: & 1.05 \:\:  & 1.40 \:\: &1.52 \:\: & 1.35 \:\: &1.67 \\
\hline
$E_d$\:\:     & -4.23 \:\:  & -4.25 \:\: & -4.23\:\: & -4.43\:\: &-4.24   \\
\hline
$(E_d-E_v)$\:\: & 0.13 \:\:  & 0.15 \:\: &0.29 \:\: & 0.44\:\: &0.50   \\

\hline
\hline
\end{tabular}
\end{table}

To further investigate the nature of the SDB states, we performed a
calculation for a Ge (110) nanowire with $D = 1.2~{\rm nm}$ in which a
GeH radical at the surface was replaced by a non-passivated Si atom,
forming a Ge wire with an SDB at a substitutional Si atom at the
surface. The band structures of this wire and of the corresponding one
for a Ge wire with a ``standard'' Ge SDB are very similar to each
other and to the one in Fig.~\ref{eeletronica}(a). The SDB state at
the Ge atom is at $E_v+~0.27~{\rm eV}$, while the SDB state at the Si
atom is at $E_v+~0.40~{\rm eV}$.  The same is observed for a Si (110)
nanowire with $D = 1.2~{\rm nm}$ in which a SiH radical at the surface
is replaced by a non-passivated Ge atom: in this case, the SDB state
at the Ge atom is at $E_v+~0.70~{\rm eV}$ and the SDB state at the Si
atom is at $E_v+~0.84~{\rm eV}$.  These two cases demonstrate that the
positions of the band edges relative to
the SDB level are weakly dependent (within 0.15 eV) on the atomic species 
where the SDB is located, but strongly dependent on the electronic structure of
the wire as a whole.

Our third numerical argument in favor of the universality of the SDB
state is based on a Ge/Si heterostructure nanowire, shown in
Fig.~\ref{axial}. We performed two separate calculations for SDB's in
this wire. In the first one, a hydrogen atom is removed from a Si
surface atom. In the second one, we removed a hydrogen atom from a Ge
surface atom. In both cases we placed the dangling bond as far as
possible from the Ge/Si interface.  Again, a striking resemblance is
observed in both band structures: the SDB level at the Ge atom is at
$E_v+~0.51$ eV while at the Si atom it is at $E_v+~0.59$ eV.

\begin{figure}[h]
\includegraphics[height=5.4cm]{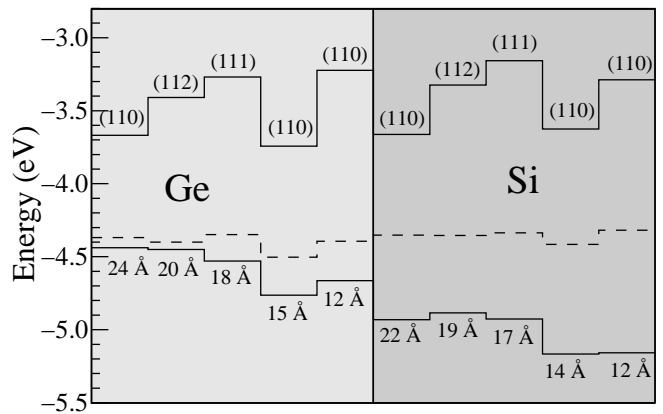}
\caption{Calculated band edges of ten Si and Ge nanowires, relative to
the vacuum level. The top line indicates the conduction band minima
and the bottom line indicates the valence band maxima.  The dashed
line corresponds to the eigenvalue of the SDB state in each case.  The
orientation and diameter of each wire is also indicated.}
\label{linha}
\end{figure}

Given the above arguments, the $\varepsilon(+/-)$ level of the
nanowire SDB's (or, likewise, its Kohn-Sham eigenvalue in GGA or LDA
approximations) should be a common reference level among Si, Ge, and
Ge/Si nanowires, and could be used, for instance, to predict
band-lineups in nanowire junctions~\cite{footnote2,hannon} and
heterostructures. This is demonstrated in Fig.~\ref{linha} where the
calculated band edges (relative to the vacuum level \cite{footnote3})
of all pure Si and Ge wires are shown together with the eigenvalues of
the SDB states.  Figure~\ref{linha} shows that the SDB
$\varepsilon(+/-)$ is a common energy level among the nanowires. The
similarity of the $\varepsilon(+/-)$ is even stronger if we remove
from the comparison the SiH$_2$- and GeH$_2$-terminated nanowires (the
ones with diameters of 1.4 and 1.5 nm, respectively). By doing that, all
$\varepsilon(+/-)$ values fall in a narrow range from $-4.40$~eV to
$-4.32$~eV.

We also performed calculations for SDB's on H-terminated Si and Ge
(111) surfaces.  We obtained $\varepsilon(+/-) = -4.41$ eV (relative
to the vacuum level) for the SDB on a Si surface, consistent with our
results for the Si nanowires. For the SDB on a Ge surface, we find a
resonant $\varepsilon(+/-)$ within the uppermost valence band,
consistent with the fact that the top of the valence band $E_v =
-$4.26~eV is above the $\varepsilon(+/-)$ nanowire values. 
In the case of the calculated Ge nanowires, Fig.~\ref{linha} shows 
that $E_v$ is always below -4.4 eV (and therefore below $\varepsilon(+/-)$),
with a a clear trend of increasing $E_v$ with diameter $D$.  
This indicates that the $E_v$ values in Ge nanowires are downshifted in energy 
relative to the surface $E_v$ value due to quantum confinement. Therefore, 
for some diameter $D$ (larger than the ones considered in this work) there will 
be a transition between localized and resonant $\varepsilon(+/-)$, with 
large-diameter Ge wires presenting a resonant $\varepsilon(+/-)$. 
This is consistent
with recent experimental results~\cite{lou} showing that
large-diameter Si-covered Ge wires are intrinsically p-type.

Why should the SDB's $\varepsilon(+/-)$ be a common energy reference?
Previously, it has been proposed that both in the case of
transition-metal impurities~\cite{tersoff2}, and of
interstitial-hydrogen impurities~\cite{naturevw03}, the common energy
reference is tied to an average dangling-bond energy which would be
independent of the host material. This proposal has never been
directly verified from first principles calculations, possibly because
isolated dangling bonds are too reactive in bulk point defects. We
argue that a surface dangling bond resulting from the removal of a
hydrogen atom from a surface GeH or SiH radical should be a good
realization of an ideal, unreconstructed dangling bond, as
the surface Si or Ge atom with an SDB is bound to
three saturated bulk-like atoms, with bond angles ($\theta$) and
lengths ($d$) that are very similar to the defectless wire values ($\Delta d <$
0.03 \AA; $\Delta \theta < 2^{\rm o}$, $6^{\rm o}$, and $1^{\rm o}$
for Si, Ge and Ge/Si wires, respectively).
Such small relaxations are indicative of small coupling between the SDB
and the underlying bulk states.

To gain an even deeper insight into the physical origin of the SDB
$\varepsilon(+/-)$ universality, let us consider this quantity in the
limit of isolated atoms. We computed the atomic-limit
$\varepsilon(+/-)$ using the same DFT approximations as in the
nanowire calculations. 
In this limit, $\varepsilon(+/-)$ is defined as the average between
the first-ionization potential and the electron affinity, 
computed for the atomic ground state.
We obtained $\varepsilon(+/-) = -$4.39~eV for Si and
$-$4.17~eV for Ge. This suggests that the universality is an atomic
property. From similar atomic calculations for group III and V
neighbors to Si and Ge, namely, Al, Ga, P, and As, we computed the
atomic-limit average ${\bar{\varepsilon}}(+/-) =
[\varepsilon(+/-)_{III} + \varepsilon(+/-)_{V}]/2$ of the III-V
compounds AlP, GaP, AlAs, and GaAs.  We obtained ${\bar{\varepsilon}}
= -$4.37~eV (AlP), $-$4.34~eV (GaP), $-$4.22~eV (AlAs), and $-$4.20~eV
(GaAs), indicating that the applicability of the SDB common energy
reference should go beyond Si and Ge.

In conclusion, {\it ab initio} calculations for hydrogen-terminated Si
and Ge nanowires, nanowire heterostructures, and Si and Ge (111)
surfaces, indicate that the charge transition levels
$\varepsilon(+/-)$ of surface dangling bond states behave as a common
energy reference among Si and Ge wires and Si/Ge
heterostructures, at 4.3 $\pm$ 0.1 eV below the vacuum
level. Calculations of $\varepsilon(+/-)$ for isolated atoms indicate
that this nearly constant value is a periodic-table atomic property.

\begin{acknowledgments}
We acknowledge support from the Brazilian agencies CNPq, FAPEMIG, and
Instituto do Mil\^enio em Nanoci\^encias-MCT.
\end{acknowledgments}


\begin{thebibliography}{100}
\bibitem{footnotejanak} This result is only valid if $\varepsilon_i$ is within the calculated
bandgap. This applies to all our nanowire calculations.
\bibitem{footnote2} Junctions made of Si nanowires with different diameters have
been recently observed~\cite{hannon}.
\bibitem{footnote3} Unlike a bulk heterojunction, a nanowire
heterojunction has a well-defined vacuum level. In the GGA or LDA
approximations, it is given by the asymptote of the Kohn-Sham
potential away from the nanowire axis, namely, E=0 . Due to the
exponential decay of $V_{KS}$ from the surface in GGA (or LDA) and to
the finite range of the basis set, the Kohn-Sham eigenvalues converge
rapidly with supercell size.
\end{thebibliography}
\end{document}